\documentclass[%
 reprint,
superscriptaddress,
 amsmath,amssymb,
 aps,
]{revtex4-2}

\usepackage{graphicx}
\usepackage{dcolumn}
\usepackage{bm}

\usepackage{xcolor}

\begin{document}

\preprint{APS/123-QED}

\title{Non-monotonic dependence of the radiation linewidth in a phonon laser}

\author{Artem R. Mukhamedyanov}
\affiliation{Moscow Institute of Physics and Technology, 141700, 9 Institutskiy pereulok, Moscow, Russia}
\author{E. S. Andrianov}
\affiliation{Moscow Institute of Physics and Technology, 141700, 9 Institutskiy pereulok, Moscow, Russia}
\affiliation{Dukhov Research Institute of Automatics (VNIIA), 127055, 22 Sushchevskaya, Moscow, Russia}
\affiliation{Institute for Theoretical and Applied Electromagnetics, 125412, 13 Izhorskaya, Moscow, Russia}
\author{A. A. Zyablovsky}
\email{zyablovskiy@mail.ru}
\affiliation{Moscow Institute of Physics and Technology, 141700, 9 Institutskiy pereulok, Moscow, Russia}
\affiliation{Dukhov Research Institute of Automatics (VNIIA), 127055, 22 Sushchevskaya, Moscow, Russia}
\affiliation{Institute for Theoretical and Applied Electromagnetics, 125412, 13 Izhorskaya, Moscow, Russia}

\date{\today}

\begin{abstract}
A phonon laser is an important device that can generate coherent sound waves at gigahertz frequency. In this paper, we demonstrate that in a phonon laser, the radiation line width can depend on the pumping non-monotonically. This is because there are two different types of solutions whose spectra depend differently on the noise amplitude. The first solution is a zero solution, which is stable before the generation threshold and is nondegenerate. The second solution is a nonzero solution that is stable above the generation threshold and is degenerate with respect to the phase. The line width of peaks in the spectrum of the zero solution does not depend on the noise amplitude. While, the line width of the peak in the spectrum of the nonzero solution increases with increasing noise amplitude. The noise amplitude increases with temperature, and, as a result, there is a temperature above which the transition from the zero solution to the nonzero solution leads to an increase in the radiation line width. In this case, the radiation line width non-monotonically depends on the pumping.
\end{abstract}

\maketitle

\section*{Introduction}
In recent decades, exciting progress has been made in the development of optomechanical systems \cite{1,2,3,4}. The efficient interaction between optical and mechanical degrees of freedom endows optomechanical systems unique properties that result from the combination of optical and acoustic vibrations. In particular, an interaction between photon and phonon modes in these systems opens the way to create phonon lasers, which are able to generate coherent sound waves with gigahertz frequency \cite{5,6,7,8,9,10,11,12,13,14,15,16}. The wavelength of such high-frequency sound waves is about a few micrometers \cite{17}, which makes them useful, for example, for imaging and manipulating micro-objects.

A phonon laser can be based on an optomechanical system consisting of several optical modes interacting with each other via phonon modes \cite{5}. In this device, the external electromagnetic wave with a frequency close to the frequency of one of the optical modes plays the role of the pumping. The generation of photons and phonons takes place when the amplitude of the external wave is above a threshold value. It is usually believed that the transition to laser generation is accompanied by a decrease in the radiation line width.

In this paper, we demonstrate that in the phonon laser, the radiation line width can non-monotonically depend on the pumping. Such non-monotonicity is observed near the generation threshold, where a zero solution ceases to be stable, and a nonzero solution becomes stable. The zero solution is nondegenerate and the line width of peaks in its spectrum of does not depend on the noise amplitude. At the same time, the nonzero solution is degenerate and therefore the line width of the peak in its spectrum depends on the noise amplitude. The noise amplitude increases with the temperature and, as a result, there is a temperature above which the transition from the zero solution to the nonzero solution leads to increase in the radiation line width. In this case, the laser line width depends non-monotonically on the pumping.

\section*{System under consideration}
We consider an optomechanical system consists of two optical modes interacting with each other via a phonon mode. To describe the considered system we use the optomechanical Hamiltonian \cite{5}:

\begin{equation}
\begin{gathered}
\hat H = \hbar {\omega _1}\hat a_1^\dag {\hat a_1} + \hbar {\omega _2}\hat a_2^\dag {\hat a_2} + \hbar {\omega _b}{\hat b^\dag }\hat b + \\
\hbar \left( {g\,\hat a_1^\dag {{\hat a}_2}\hat b + {g^*}\,{{\hat a}_1}\hat a_2^\dag {{\hat b}^\dag }} \right) + 
\hbar \left( {\Omega \,\hat a_1^\dag {e^{ - i{\omega _1}\,t}} + {\Omega ^*}{{\hat a}_1}{e^{i{\omega _1}\,t}}} \right)
\label{eq:1}
\end{gathered}
\end{equation}

Here ${\hat a_{1,2}}$ and $\hat a_{1,2}^\dag $ are the annihilation and creation bosonic operators for the first and the second optical modes, respectively. $\hat b$ and ${\hat b^\dag }$ are the annihilation and creation operators of a phonon mode. ${\omega _{1,2}}$ are frequencies of the optical modes. ${\omega _b}$ is a frequency of the phonon mode. $g$ is a coupling strength between the modes and the phonons (Frohlich constant). $\Omega $ is an amplitude of the external wave; a frequency of the external wave is equal to ${\omega _1}$ that corresponds to a resonant excitation of the system.

Within the framework of Heisenberg-Langevin approach \cite{18,19}, we derive the equations for the expected values of amplitudes of the optical modes and phonon mode \cite{16}:

\begin{equation}
\frac{{d{a_1}}}{{dt}} = \left( { - {\gamma _1} - i{\omega _1}} \right){a_1} - ig{a_2}b - i\Omega {e^{ - i{\omega _1}\,t}}
\label{eq:2}
\end{equation}

\begin{equation}
\frac{{d{a_2}}}{{dt}} = \left( { - {\gamma _2} - i{\omega _2}} \right){a_2} - i{g^*}{a_1}{b^*}
\label{eq:3}
\end{equation}

\begin{equation}
\frac{{db}}{{dt}} = \left( { - {\gamma _b} - i{\omega _b}} \right)b - i{g^*}{a_1}a_2^* + \xi \left( t \right)
\label{eq:4}
\end{equation}

Here ${a_{1,2}} = \left\langle {{{\hat a}_{1,2}}} \right\rangle $, $b = \left\langle {\hat b} \right\rangle $ are the expected values of amplitudes of the optical modes and phonon mode, respectively. ${\gamma _{1,2}}$, ${\gamma _b}$ are the relaxation rates of the respective quantities. $\xi \left( t \right)$ is a thermal noise acting on the phonon mode, the correlation function of which is proportional to ${\gamma _b}\bar n$, where $\bar n$ is a average number of thermal phonons in the system. We consider that the system temperature is much lower than the frequencies of the optical mode ($kT <  < \hbar {\omega _{1,2}}$) and, therefore, we neglect a noise acting on the optical modes.

\section*{Stationary states}
To start, we study the system behavior without taking into account noise. We are looking for a stationary solution of Eqns.~(\ref{eq:2})-(\ref{eq:4}) (by taking $\xi  = 0$) in the form ${a_1} = {a_{1st}}{e^{ - i{\omega _1}\,t}}$, ${a_2} = {a_{2st}}{e^{ - i\left( {{\omega _1} - \delta \omega } \right)t}}$, $b = {b_{st}}{e^{ - i\delta \omega t}}$, where ${a_{1st}}$, ${a_{2st}}$, ${b_{st}}$ do not depend on time and $\delta \omega$ is a frequency of generated phonons \cite{16}. In the considered system, there is a zero solution, for which ${a_{2st}} = {b_{st}} = 0$, ${a_{1st}} =  - i\,\Omega /{\gamma _1}$ \cite{16} that corresponds to linear excitation of the first mode by the external field. A linear analysis of stability of the zero solution shows that small deviations from the zero solution are obtained by the following linear equations:

\begin{equation}
\frac{d}{{dt}}\left( {\begin{array}{*{20}{c}}
  {\delta {a_1}} \\ 
  {\delta {a_2}} \\ 
  {\delta b} 
\end{array}} \right) = \hat M\left( {\begin{array}{*{20}{c}}
  {\delta {a_1}} \\ 
  {\delta {a_2}} \\ 
  {\delta b} 
\end{array}} \right)
\label{eq:5}
\end{equation}
Here 
\begin{equation}
\hat M = \left( {\begin{array}{*{20}{c}}
  { - {\gamma _1}}&0&0 \\ 
  0&{ - i\left( {\delta {\omega _2} + \delta \omega } \right) - {\gamma _2}}&{ - {g^*}\,\Omega /{\gamma _1}} \\ 
  0&{ - g\,{\Omega ^*}/{\gamma _1}}&{i\left( {{\omega _b} - \delta \omega } \right) - {\gamma _b}} 
\end{array}} \right)
\label{eq:6}
\end{equation}
where $\delta {\omega _2} = {\omega _2} - {\omega _1}$.

The eigenvalues of Eqns.~(\ref{eq:5}), (\ref{eq:6}) are given as
\begin{equation}
{\lambda _1} =  - {\gamma _1}
\label{eq:7}
\end{equation}

\begin{equation}
\begin{gathered}
{\lambda _{2,3}} =  - \frac{{{\gamma _2} + {\gamma _b} + i\left( {\left( {\delta {\omega _2} + \delta \omega } \right) - \left( {{\omega _b} - \delta \omega } \right)} \right)}}{2} \\
\pm \frac{1}{2}\sqrt {{{\left( {\left( {{\gamma _2} - {\gamma _b}} \right) + i\left( {\delta {\omega _2} + {\omega _b}} \right)} \right)}^2} + \frac{{4{{\left| g \right|}^2}{{\left| \Omega  \right|}^2}}}{{\gamma _1^2}}}
\label{eq:8}
\end{gathered}
\end{equation}

The real parts of all eigenvalues are less than zero when $\Omega  < {\Omega _{th}} = \frac{{{\gamma _1}\,{\gamma _2}}}{{\left| g \right|}}\sqrt {\frac{{{\gamma _b}}}{{{\gamma _2}}}} \sqrt {1 + {{\left( {\frac{{\delta {\omega _2} + {\omega _b}}}{{{\gamma _2} + {\gamma _b}}}} \right)}^2}} $. That is, the zero solution is stable when $\Omega  < {\Omega _{th}}$.

Besides the zero solution, there is a nonzero solution of Eqns.~(\ref{eq:2})-(\ref{eq:4}) without noise, for which the stationary values of the amplitudes of the optical modes and the phonon mode are given as

\begin{equation}
{\left| {{a_{1st}}} \right|^2} = \frac{{\Omega _{th}^2}}{{\gamma _1^2}}
\label{eq:9}
\end{equation}

\begin{equation}
{\left| {{a_{2st}}} \right|^2} = \frac{1}{{\left| g \right|}}\sqrt {\frac{{{\gamma _b}}}{{{\gamma _2}}}} \sqrt {{{\left| \Omega  \right|}^2} - \Omega _{ex}^2}  - \frac{{{\gamma _1}{\gamma _b}}}{{{{\left| g \right|}^2}}}
\label{eq:10}
\end{equation}

\begin{equation}
{\left| {{b_{st}}} \right|^2} = \frac{1}{{\left| g \right|}}\sqrt {\frac{{{\gamma _2}}}{{{\gamma _b}}}} \sqrt {{{\left| \Omega  \right|}^2} - \Omega _{ex}^2}  - \frac{{{\gamma _1}{\gamma _2}}}{{{{\left| g \right|}^2}}}
\label{eq:11}
\end{equation}

Here ${\Omega _{ex}} = \frac{{{\gamma _1}{\gamma _2}}}{{\left| g \right|}}\sqrt {\frac{{{\gamma _b}}}{{{\gamma _2}}}} \frac{{\delta {\omega _2} + {\omega _b}}}{{{\gamma _2} + {\gamma _b}}}$. The nonzero solutions exist when $\Omega  \geqslant {\Omega _{th}}$ (i.e., ${\left| {{a_{2st}}} \right|^2} > 0$ and ${\left| {{b_{st}}} \right|^2} > 0$) \cite{16}. It is important that the nonzero solution is degenerate with respect to the phase difference between the second optical mode and the phonon mode. That is, there are an infinite number of nonzero solutions differing from each other only in the phase difference.

To determine the external wave amplitude for which the nonzero solution is stable, we make a linear analyze of stability. It is determined that there is an eigenvalue whose real part is equal to zero.
This reflects the phase degeneracy of non-zero solutions. The real parts of other eigenvalues are less than zero when $\Omega  \geqslant {\Omega _{ex}}$. Thus, the family of nonzero solutions is stable when $\Omega  \geqslant {\Omega _{ex}}$ \cite{16}.

The transition to lasing occurs when the system switches from the zero solution to the nonzero solution. This transition is accompanied by a change in the laser spectrum (in particular, a change in the line width). Further, we show that the frequency spectra of the modes depend differently on the external wave amplitude and noise amplitudes when the system evolves near the zero solution and near the non-zero solution, leading to non-monotonic line width behavior.

\section*{Spectra for zero and nonzero states}
\subsection*{Spectrum of the zero solution}
\begin{figure}[htbp]
\centering\includegraphics[width=\linewidth]{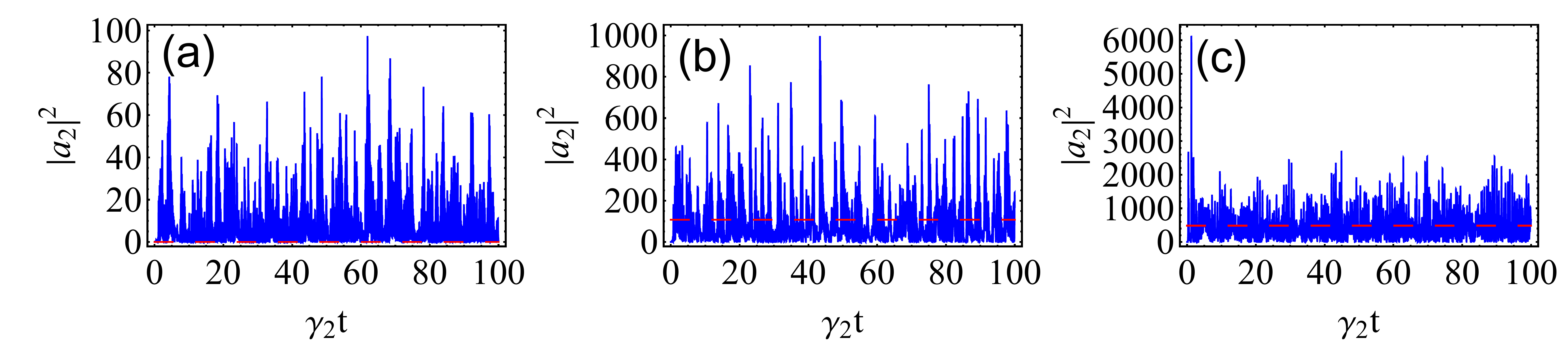}
\caption{Temporal dependence of the intensity of the second optical mode. (a) $\Omega  = 5 \cdot {10^{ - 3}}\,{\omega _0} < {\Omega _{th}}$; (b) $\Omega  = 1.5 \cdot {10^{ - 2}}\,{\omega _0} > {\Omega _{th}}$; (c) $\Omega  = 5 \cdot {10^{ - 2}}{\omega _0} >  > {\Omega _{th}}$. The other parameters are ${\gamma _1} = {\gamma _2} = {\gamma _b} = 2 \cdot {10^{ - 4}}{\omega _0}$, $\delta {\omega _1} = 0$, $\delta {\omega _2} = 5 \cdot {10^{ - 3}}{\omega _0}$, ${\omega _b} = 5 \cdot {10^{ - 3}}{\omega _0}$, $g = {10^{ - 4}}{\omega _0}$, and $\bar n = {10^2}$. At these parameters ${\Omega _{th}} = {10^{ - 2}}\,{\omega _0}$.}
\label{fig1}
\end{figure}

\begin{figure}[htbp]
\centering\includegraphics[width=\linewidth]{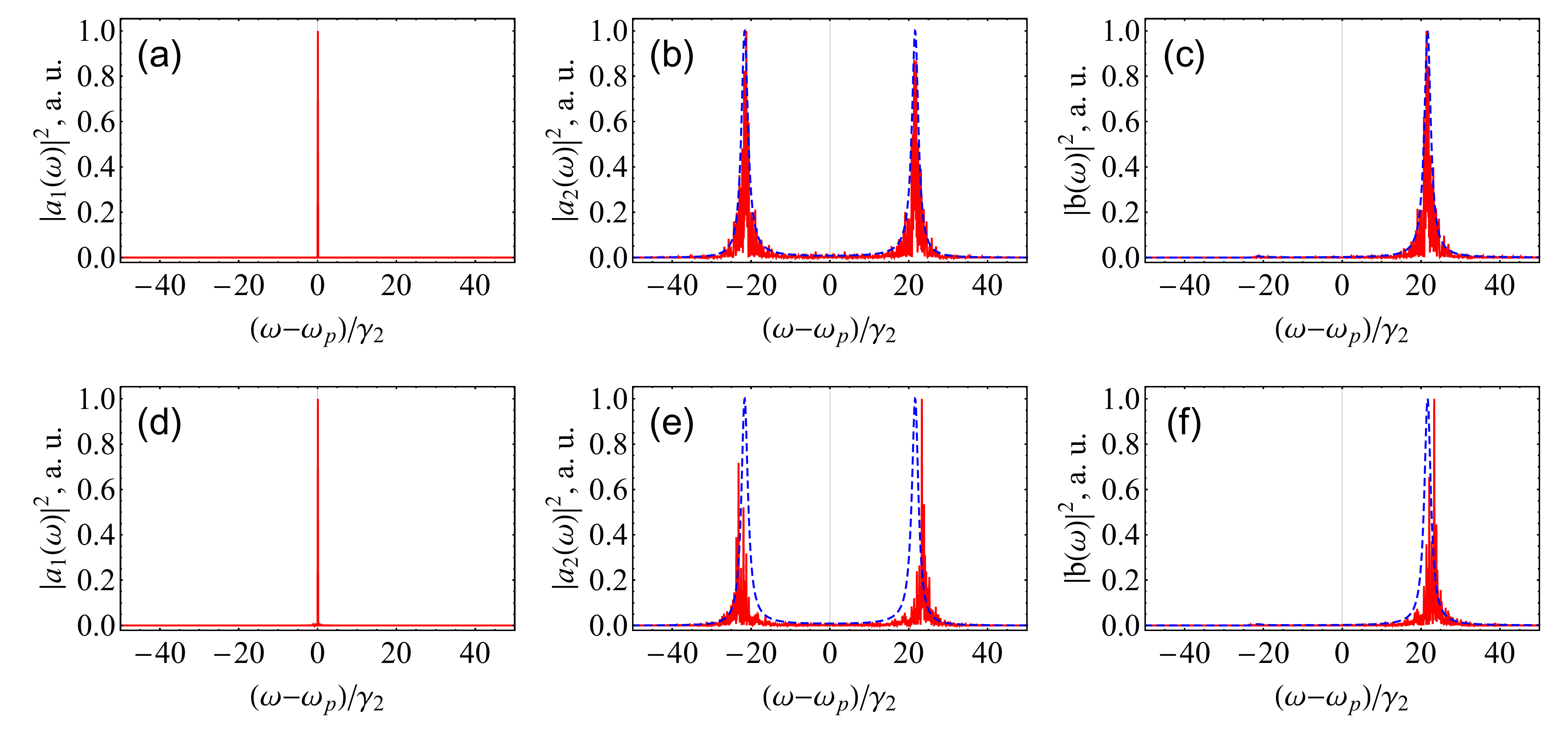}
\caption{Spectra of the first optical mode (a), (d), the second optical mode (b), (d) and the phonon mode (c), (e) when the system evolves near the zero solution. Solid lines are obtained by numerical modeling of Eqns.~(\ref{eq:2})-(\ref{eq:4}). Dashed lines are obtained by formula~\ref{eq:12}. Here ${\gamma _1} = {\gamma _2} = {\gamma _b} = 2 \cdot {10^{ - 4}}{\omega _0}$, $\delta {\omega _{\,1}} = 0$, $\delta {\omega _2} = 5 \cdot {10^{ - 3}}{\omega _0}$, ${\omega _b} = 5 \cdot {10^{ - 3}}{\omega _0}$, $g = {10^{ - 4}}{\omega _0}$, $\Omega  = 5 \cdot {10^{ - 3}}{\omega _0}$. (a), (b), (c) $\bar n = 10$; (d), (e), (f) $\bar n = {10^2}$.}
\label{fig2}
\end{figure}

In the considered system, the noise leads to fluctuations near the stable states [Figure~\ref{fig1}]. When the system evolves near the zero state, the noise completely determines the amplitudes of the second optical mode and the phonon mode. Numerical simulations of Eqns.~(\ref{eq:2})-(\ref{eq:4}) show that as long as the deviations from the steady state caused by the noise remain small, the shape of the spectrum for the corresponding modes is independent of the noise amplitude [Figure~\ref{fig2}]. The spectrum of the first optical mode has a sharp peak at ${\omega _1}$ [Figure~\ref{fig2}a]. The spectrum of the second optical mode has two symmetric peaks [Figure~\ref{fig2}b]. As the amplitude of the external wave, $\Omega $, increases, the distance between the peaks decreases. The spectrum of the phonon mode has a peak with a larger amplitude at $\omega  > {\omega _1}$ and a peak with a smaller amplitude at $\omega  < {\omega _1}$ [Figure ~\ref{fig2}b].

The mode spectra can also be calculated analytically using the expression \cite{18}:

\begin{equation}
\hat S\left( \omega  \right) = \frac{1}{\pi }{\left( {{{\hat M}^*} + i\omega \hat I} \right)^{ - 1}}\hat D{\left( {{{\hat M}^T} - i\omega \hat I} \right)^{ - 1}}
\label{eq:12}
\end{equation}

Here $\hat S\left( \omega  \right)$ is the matrix whose diagonal elements define the spectra of each of the modes, $\hat I$ is a three-by-three identity matrix, and

\begin{equation}
\hat D = \left( {\begin{array}{*{20}{c}}
  0&0&0 \\ 
  0&0&0 \\ 
  0&0&{{\gamma _b}\bar n} 
\end{array}} \right)
\label{eq:13}
\end{equation}
is the diffusion matrix of the noise in Eqns.~(\ref{eq:2})-(\ref{eq:4}).

The spectra obtained by numerical modeling of Eqns.~(\ref{eq:2})-(\ref{eq:4}) are in good agreement with the spectra found using Eq.~(\ref{eq:12}) [Figure~\ref{fig2}].

Expression~(\ref{eq:12}) is obtained under the assumption that noise leads to small fluctuations near the steady state. Under this assumption, the noise amplitude, which is proportional to ${{\gamma _b}\bar n}$, does not affect the type of diffusion matrix and, as a consequence, the shape of the spectrum, but only determines the common factor in the expression for the spectrum (see Eq.~(\ref{eq:12})). That is, increasing noise only leads to an increase in the amplitude of the spectrum. Since the intensities of the second optical and phonon modes are determined exclusively by noise, the laser signal is incoherent.

\subsection*{Spectrum of the nonzero solution}
\begin{figure}[htbp]
\centering\includegraphics[width=\linewidth]{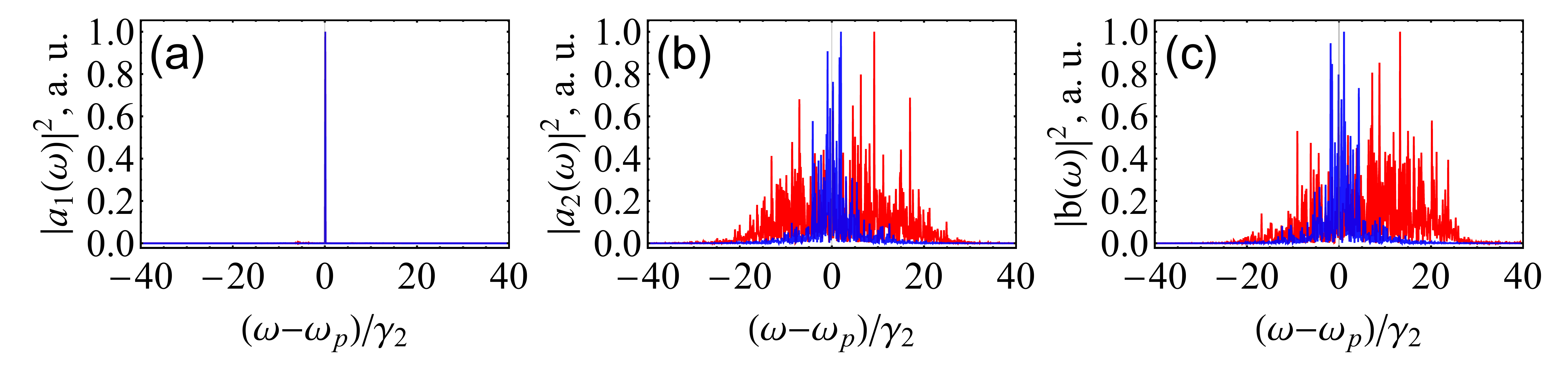}
\caption{Spectra of the first optical mode (a), the second optical mode (b) and the phonon mode (c) when the system evolves near the nonzero solution. Blue lines are obtained at $\bar n = 10$; red lines at $\bar n = {10^2}$. Here ${\gamma _1} = {\gamma _2} = {\gamma _b} = 2 \cdot {10^{ - 4}}{\omega _0}$, $\delta {\omega _{\,1}} = 0$, $\delta {\omega _2} = 5 \cdot {10^{ - 3}}{\omega _0}$, ${\omega _b} = 5 \cdot {10^{ - 3}}{\omega _0}$, $g = {10^{ - 4}}{\omega _0}$, $\Omega  = 3 \cdot {10^{ - 2}}{\omega _0}$.}
\label{fig3}
\end{figure}

When the system evolves near the nonzero stationary solution, the noise leads to smaller relative fluctuations in intensity than in the case of the zero solution. The spectra of all modes have one maximum at $\omega  = {\omega _1}$ [Figure~\ref{fig3}].

Since the nonzero solution is degenerate with respect to phase difference, the noise leads to a random walk between the states with different phases. Such a walk occurs along a direction in the space of vectors $\left( {\delta {a_1},\,\delta a_1^*,\,\delta {a_2},\,\delta a_2^*,\delta b,\,\delta {b^*}} \right)$, which is determined by the eigenvector with the zero eigenvalue. The magnitude of the walk depends on the noise amplitude, so the mode spectra also depend on the noise amplitude [Figure~\ref{fig3}] \cite{18}.

\begin{figure}[htbp]
\centering\includegraphics[width=0.5\linewidth]{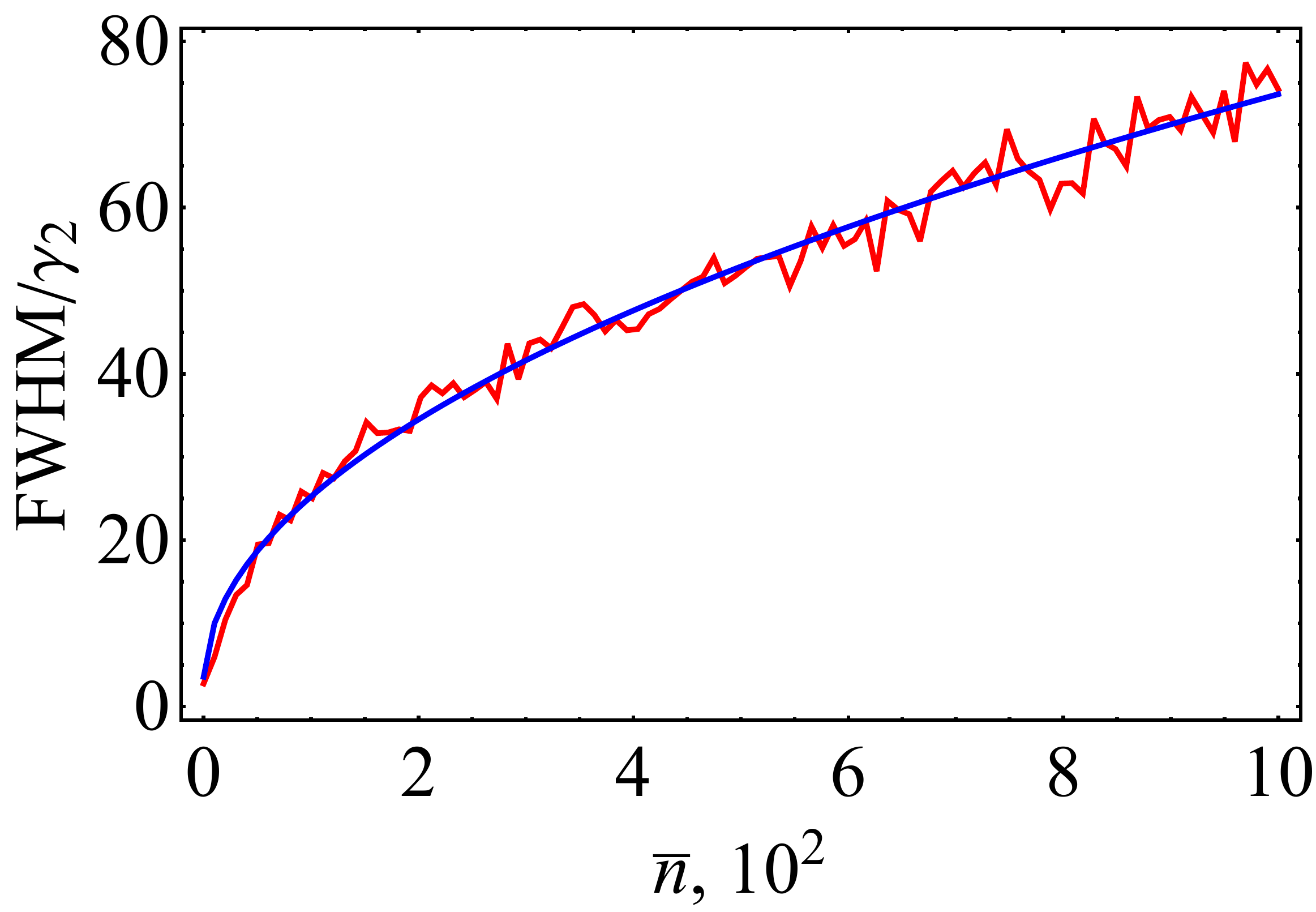}
\caption{Dependence of the line width in the spectrum of the second optical mode on $\bar n$, when the system evolves near the nonzero solutions. Here ${\gamma _1} = {\gamma _2} = {\gamma _b} = 2 \cdot {10^{ - 4}}{\omega _0}$, $\delta {\omega _{\,1}} = 0$, $\delta {\omega _2} = 5 \cdot {10^{ - 3}}{\omega _0}$, $\delta {\omega _b} = 5 \cdot {10^{ - 3}}{\omega _0}$, $g = {10^{ - 4}}{\omega _0}$, $\Omega  = 5 \cdot {10^{ - 2}}{\omega _0}$.}
\label{fig4}
\end{figure}

To determine the dependence of the line width on the noise amplitude, we performed numerical simulations of Eqns.~(\ref{eq:2})-(\ref{eq:4}) for different values of $\bar n$. Our calculations show that the line width in the spectrum of the second optical mode and the phonon mode is proportional to ${\bar n^{1/2}}$ [Figure~\ref{fig4}].

\begin{figure}[htbp]
\centering\includegraphics[width=\linewidth]{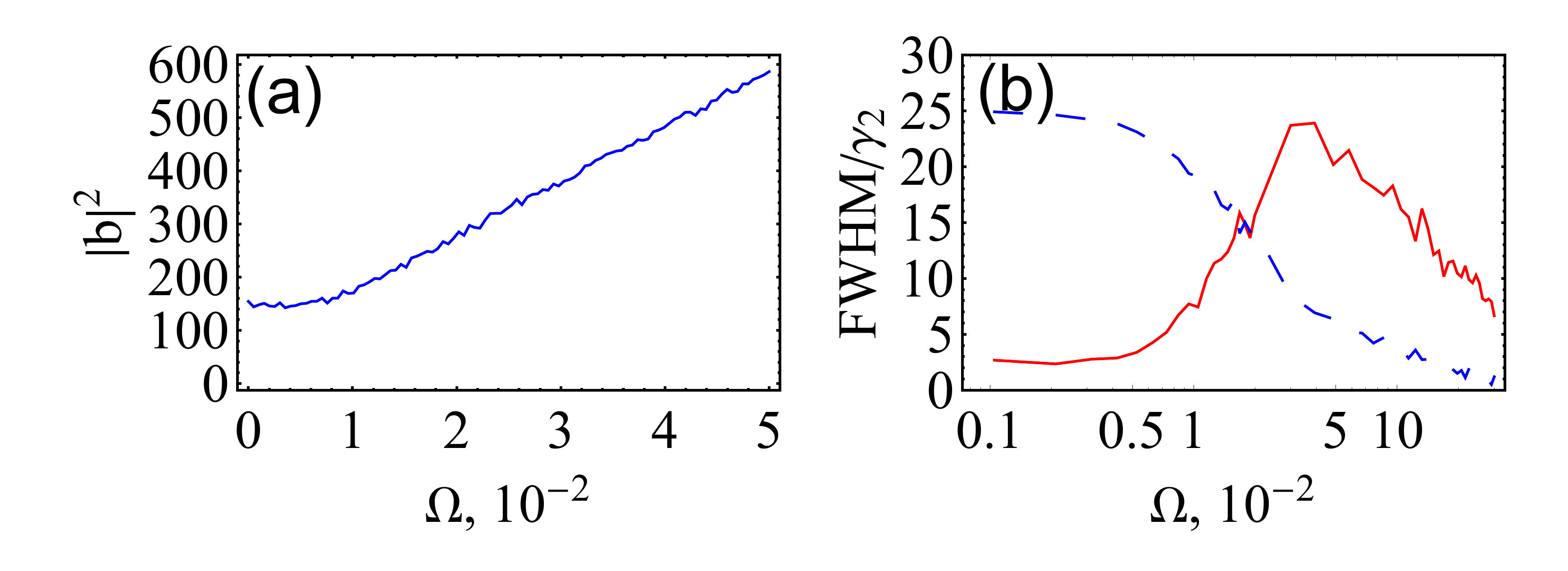}
\caption{(a) Dependence of the intensity of the phonon mode on the amplitude of the external wave. (b) Dependence of the line width (the solid red line) and the peak frequency (the dashed blue line) of the phonon mode on the amplitude of the external wave. Here $\bar n = {10^2}$, ${\gamma _1} = {\gamma _2} = {\gamma _b} = 2 \cdot {10^{ - 4}}{\omega _0}$, $\delta {\omega _{\,1}} = 0$, $\delta {\omega _2} = 5 \cdot {10^{ - 3}}{\omega _0}$, ${\omega _b} = 5 \cdot {10^{ - 3}}{\omega _0}$, $g = {10^{ - 4}}{\omega _0}$. At these parameters ${\Omega _{th}} \approx {10^{ - 2}}{\omega _0}$.}
\label{fig5}
\end{figure}

Thus, we have shown that the line width in the spectrum of the zero solution does not depend on the noise amplitude. At the same time, the line width in the spectrum of the nonzero solution increases with the noise amplitude. Therefore, there is a critical value of the noise amplitude above which the line width of the nonzero solution exceeds the line width of the zero solution.

The transition to laser generation corresponds to the switching of the system from the zero solution to the non-zero one. Below the lasing threshold, the spectrum of the system is determined by the spectrum near the zero solution. At small values of the external wave amplitude, the spectrum contains two peaks [Figure~\ref{fig2}]. As the external wave amplitude increases, the distance between the peaks decreases and tends to zero near the threshold. Above the lasing threshold, the spectrum contains one peak. If the noise amplitude is greater than the critical value, the width of the peaks depends non-monotonically on the external wave amplitude [Figure~\ref{fig5}b]. The line width increases near the lasing threshold because the line width of the zero solution does not depend on the noise intensity, and the line width of the nonzero solution depends on the noise intensity. With a further increase in the pump wave amplitude above the lasing threshold, the line width decreases, which is standard behavior for lasing.

\section*{Conclusion}
To summarize, we have shown that in a phonon laser, the line widths of the generated photons and phonons may not depend monotonically on the amplitude of the external pump wave. There is a critical value of the noise amplitude (temperature), above which the line width below the generation threshold is smaller than the line width after the generation threshold. In this case, a narrow spectral line corresponds to low radiation intensity, and a wider radiation line corresponds to a high intensity. This behavior is due to the fact that the nonzero solution, which is realized after the threshold, is degenerate in the phase difference between the optical and phonon modes. Noise leads to a random walk between degenerate states that results in the dependence of the line width on the noise amplitude. In turn, the line width for the zero solution, which is stable below the generation threshold, does not depend on the noise amplitude. This fact is the reason for the existence of a critical value for the noise amplitude (temperature). The obtained result can be useful for designing sources of coherent radiation based on optomechanical systems.

\section*{Funding}
Russian Science Foundation (No. 20-72-10057).

\section*{Acknowledgments}
The study was financially supported by a Grant from Russian Science Foundation (project No. 20-72-10057). A.R.M. and E.S.A. thank the foundation for the advancement of theoretical physics and mathematics “Basis”.

\nocite{*}

\bibliography{apssamp}

\end{document}